\documentclass{emulateapj}
\usepackage{psfig}

\newcommand{\RA}[3]{{#1}^{{\rm h}}{#2}^{{\rm m}}{#3}^{{\rm s}}}
\newcommand{\Dec}[3]{{#1}^{\circ}{#2}'{#3}''}
\newcommand{\E}[1]{\times 10^{#1}}

      \newcommand{\ps}{\,{\rm s}^{-1}}
    \newcommand{\Msun}{M_{\odot}}   \newcommand{\Zsun}{Z_{\odot}}
\newcommand{\cm}{\,{\rm cm}}    \newcommand{\km}{\,{\rm km}}
\newcommand{\kpc}{\,{\rm kpc}}
        
    \newcommand{\keV}{\,{\rm keV}}

\newcommand{\nX}{n_{\rm X}}     \newcommand{\du}{d_{6.2}}
\newcommand{\HH}{H$_2$}
\newcommand{\VLSR}{V_{\rm LSR}}
\newcommand{\twCO}{$^{12}$CO}  \newcommand{\thCO}{$^{13}$CO}
\newcommand{\RCO}{$^{12}$CO $J$= 2--1/$J$= 1--0}

\begin{document}

\title{Molecular environment and thermal X-ray spectroscopy
of the semicircular young composite supernova remnant 3C~396}

\shorttitle{Molecular environment and X-rays of SNR 3C~396}

\author{
Yang Su\altaffilmark{1,2}, Yang
Chen\altaffilmark{2,3,}\footnotemark[7], 
Ji
Yang\altaffilmark{1,4}, Bon-Chul Koo\altaffilmark{5}, Xin
Zhou\altaffilmark{2}, Deng-Rong Lu\altaffilmark{1}, Il-Gyo
Jeong\altaffilmark{5}, and Tracey DeLaney\altaffilmark{6}}

\affil{ $^1$ Purple Mountain Observatory, Chinese Academy of
Sciences, Nanjing 210008, China \\
$^2$ Department of Astronomy, Nanjing
University, Nanjing 210093, China \\
$^3$ Key Laboratory of Modern Astronomy and Astrophysics,
Nanjing University, Ministry of Education, Nanjing 210093, China \\
$^4$ Key Laboratory of Radio Astronomy, Chinese Academy of
Sciences, Nanjing 210008, China \\
$^5$ Astronomy Program, Department of Physics and Astronomy, Seoul
National University, Seoul 151-747, Korea \\
$^6$ Kavli Institute for Astrophysics and Space Research,
Massachusetts Institute of Technology, 70 Vassar Street,
Cambridge, MA 02139, USA
}

\footnotetext[7]{Corresponding author to whom any correspondence
should be addressed.}

\begin{abstract}
We have investigated the molecular environment of the semicircular
composite supernova remnant (SNR) 3C~396 and
performed a {\sl Chandra} spatially resolved thermal X-ray spectroscopic
study of this young SNR.
With our CO millimeter observations, we find that
the molecular clouds (MCs) at $V_{\rm LSR}\sim$~84~km~s$^{-1}$
can better explain the multiwavelength properties of the remnant
than the $V_{\rm LSR}$= 67--72~km~s$^{-1}$ MCs that are
suggested by Lee et al.\ (2009).
At around $84\km\ps$, the western boundary of the SNR is perfectly
confined by the western molecular wall.
The CO emission fades out from
west to east, indicating that the eastern region is of low gas
density.
In particular, an intruding finger/pillar-like MC,
which may be shocked at the tip, can well explain the X-ray and radio
enhancement in the southwest and some infrared filaments there.
The SNR--MC interaction is also favored
by the relatively elevated \twCO\ $J$=2--1/$J$=1--0 line ratios in the
southwestern ``pillar tip" and the molecular patch on the northwestern
boundary. The redshifted \twCO\ ($J$=1--0 and $J$=2--1) wings
(86--90~km~s$^{-1}$) of an eastern $81\km\ps$
molecular patch may be the kinematic evidence for shock--MC
interaction.
We suggest that the 69~km~s$^{-1}$ MCs are in the
foreground based on HI self-absorption while the 84~km~s$^{-1}$ MCs
at a distance of 6.2 kpc (the tangent point) are in physical contact with SNR 3C~396.
The X-ray spectral analysis suggests an SNR age of $\sim3$~kyr.
The metal enrichment of the X-ray emitting gas in the north and south
implies a 13--$15\Msun$ B1--B2 progenitor star.

\end{abstract}


\section{INTRODUCTION}

The progenitors of core-collapse supernovae are massive stars and
they most likely formed in giant molecular clouds (MCs). Because
of their short lifetime, one may expect that the supernova
remnants (SNRs) of core-collapse supernovae would be near their
natal environment. The association
with MCs often makes SNRs display irregular morphologies in
multiwavelengths, which indicates complicated shock interaction
with the complex, inhomogeneous ambient interstellar medium (ISM).

3C~396 (G39.2$-$0.3) is a semicircular young composite SNR, exhibiting
both the central X-ray bright pulsar wind nebula (PWN) and western
incomplete half-shell structures in multiwavelengths. In the high
dynamic-range VLA images of the remnant, a remarkable ``tail''
extending from the very faint eastern shell of the SNR is revealed
and is interpreted as an eastern ``blow-out'' structure (Patnaik
et al.\ 1990). From VLA 6 and 20~cm data, a large scale brightness
asymmetry is found in the east--west direction across SNR 3C~396,
the western half of the remnant having roughly twice the
flux density of the east (Anderson \& Rudnick 1993). The {\sl
Einstein} X-ray emission is bright in the limb and the center of
the remnant (Becker \& Helfand 1987). Based on {\sl ASCA} X-ray
spectral analysis, a nonthermal component in the center of 3C~396
is attributed to a synchrotron PWN (Hurrus \& Slane 1999).
Recently, with about 100~ks high spatial resolution {\sl Chandra}
observation, a nonthermal nebula containing a point-like source is
also found in the center of the remnant and is
interpreted as a synchrotron PWN surrounding a yet undetected
pulsar (Olbert et al. 2003). The possibility of the excess
microwave emission arising from a spinning dust component toward
3C~396 is investigated at 33~GHz with the Very Small Array (VSA)
telescope (Scaife et al.\ 2007). Based on the detection of H$_2$
molecules and the sharp southwestern boundary in the radio
emission, the remnant is suggested to be encountering a
MC in this region (Lee et al.\ 2009). The remnant is
suggested at a distance of about 8.5~kpc because a cavity-like
structure of the $V_{\rm LSR}\sim$~69~km~s$^{-1}$ molecular gas
seems to surround the remnant. The {\sl Spitzer} IRS
observation reveals that, along the western shell, the inner
[FeII]-line filament and the outer H$_2$-line one is found to be spatially
separated
(Hewitt et al.\ 2009), which is similar to the result of near-IR
1.64~$\mu$m [FeII] and 2.12~$\mu$m H$_2$ line observations using
Wide-field Infrared Camera aboard the Palomar 5 m Hale telescope
(Lee et al.\ 2009). It is most probably because of the various
molecular gas density conditions of the multi-phase preshock ISM
across the western shell of 3C~396, where the shock of the remnant
is interacting with a complicated distribution of MCs (Hewitt et
al.\ 2009). A detailed investigation of the molecular gas
environment and the hot gas interaction with it are thus necessary,
which may be helpful to understand the multiwavelength physical
properties and the dynamic evolution of this irregularly shaped
SNR.

In this paper, we have presented our millimeter CO line observations
toward SNR 3C~396 and a spatially resolved {\sl Chandra} thermal X-ray
analysis.
The observations and the data
reduction are described in Section 2. The main observational
results and calculations are presented in Section 3. The
discussion and conclusions are given in Section 4 and Section 5,
respectively.

\section{OBSERVATIONS AND DATA REDUCTION}

The observation was first made in the \twCO~($J$=1--0) line (at
115.271~GHz) in 2006 April using the 6 m millimeter-wavelength
telescope of the Seoul Radio Astronomy Observatory (SRAO) with
single-side band filter and the 1024-channel autocorrelator with
50 MHz bandwidth. The half-power beam size at 115~GHz is $2'$ and
the typical rms noise level was about 0.2~K at the 0.5~km~s$^{-1}$
velocity resolution. The radiation temperature is determined by
$T_{\rm R}=T_{\rm A}/(f_{\rm b}\times\eta_{\rm mb})$, where
$T_{\rm A}$ is the antenna temperature, $f_{\rm b}$ the beam
filling factor (assuming $f_{\rm b}\sim$~1), and $\eta_{\rm mb}$
the main beam efficiency ($\sim$~75$\%$). We mapped the $14'
\times 15'$ area covering 3C~396 centered at
$(\RA{19}{04}{06},\Dec{05}{27}{00})$ with grid spacing of $\sim$1$'$.

The follow-up observations were made simultaneously in the
\twCO~($J$=1--0) line (at 115.271~GHz), \thCO~($J$=1--0) line
(110.201~GHz), and C$^{18}$O~($J$=1--0) line (109.782~GHz) during
2008 November--December using the 13.7 m millimeter-wavelength
telescope of the Purple Mountain Observatory at Delingha
(hereafter PMOD). The spectrometer has 1024 channels, with a total
bandwidth of 145~MHz (43~MHz) and the velocity resolution of
0.5~km~s$^{-1}$ (0.2~km~s$^{-1}$) for \twCO ($^{13}$CO and
C$^{18}$O). The mean rms noise level of antenna temperature was
less than 0.1~K for those lines. The intensity of the
C$^{18}$O~($J$=1--0) emission is too weak to examine its spatial
distribution. The half-power beamwidth of the telescope is about
$54''$ and the main beam efficiency $\eta_{\rm mb}$ was 52$\%$. We
mapped the $9' \times 9'.5$ area centered at
$(\RA{19}{04}{03},\Dec{05}{26}{30})$ with grid spacing of $30''$.

We also made the \twCO~($J$=2--1) line observation toward partial
regions of the SNR with grid spacing of $30''$ in 2010
February using the telescope of the SRAO to complement the data.
The half-power beam size at 230~GHz is
$48''$, the main beam efficiency was $\sim$~57$\%$, and the typical
rms noise level was about 0.2~K at the 0.13~km~s$^{-1}$ velocity
resolution.
The resolution of the SRAO \twCO~($J$=2--1) observation
and the PMOD \twCO~($J$=1--0) observation are very similar,
which makes it convenient to directly compare the two observations.

All the CO observation data were reduced using the GILDAS/CLASS
package developed by
IRAM.\footnote{http://www.iram.fr/IRAMFR/GILDAS}

The {\sl Chandra} X-ray and {\sl Spitzer} 24~$\mu$m mid-IR
observations are also used for analysis. We revisited the {\sl
Chandra} ACIS observational data of SNR 3C~396 (ObsIDs: 1988; PI:
S.~P.\ Reynolds) with total exposure time of about 100~ks. We
reprocessed the event files (from Level 1 to Level 2) using the
CIAO3.4 data processing software to remove pixel randomization and
to correct for CCD charge-transfer inefficiencies. The overall
light curve was examined for possible contamination from a
time-variable background. The reduced data, with a net exposure of
93.7~ks, were used for our subsequent analysis. The mid-IR
24~$\mu$m observation used here was carried out as the 24 and 70
Micron Survey of the Inner Galactic Disk Program (PID: 20597; PI:
S.\ Carey) with the Multiband Imaging Photometer (MIPS) (Rieke et
al.\ 2004). The Post Basic Calibrated Data (PBCD) of the 24~$\mu$m
mid-IR were obtained directly from {\sl Spitzer} archive. The VLA
1.4~GHz radio continuum emission (L band) data were adopted from
Anderson \& Rudnick (1993). The HI-line data and the large scale \thCO\ line data
were obtained from the archival VLA Galactic Plane Survey (VGPS: Stil et al.\ 2006)
and the Galactic Ring Survey (GRS: Jackson et al.\ 2006), respectively.

\section{RESULTS}
\subsection{Molecular Gas Environment of SNR 3C~396}
\label{HICO}

\begin{figure*}[tbh!]
\centerline{ {\hfil\hfil
\psfig{figure=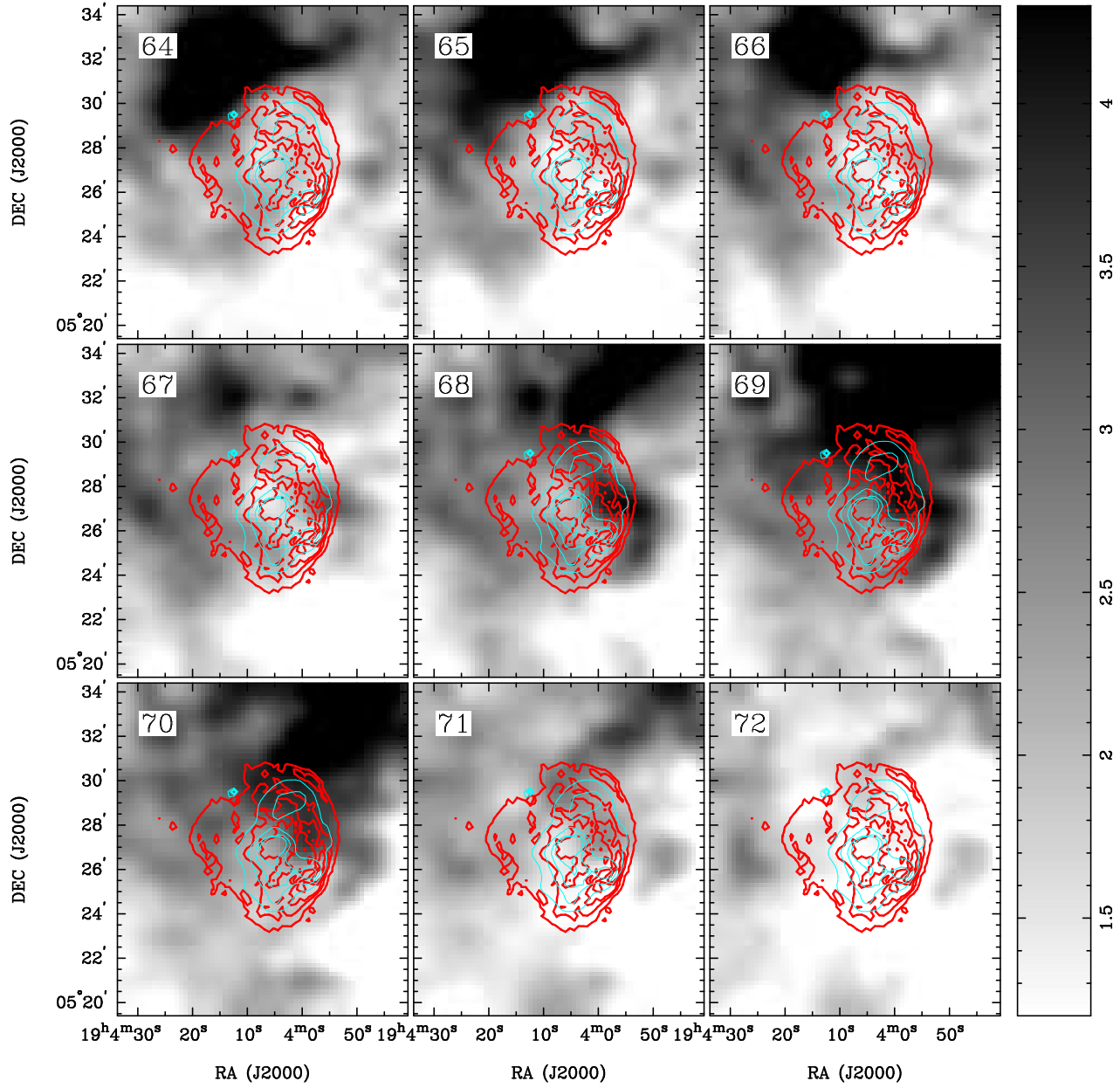,height=6.5in,angle=270, clip=} \hfil\hfil}}
\caption{SRAO \twCO\ ($J$=1--0) intensity channel maps between 64
and 72 km~s$^{-1}$ (smoothed to a resolution of $0'.2$ by
interpolation), overlaid with the 1.0--7.0~keV X-ray ({\em cyan,
thin}) contours and the 1.4~GHz radio ({\em red, thick}) continuum
emission contours at levels of 1.5, 3.5, 5.5, 7.5, 9.5 mJy
beam$^{-1}$.} \label{f:srao6472}
\end{figure*}

\begin{figure*}[tbh!]
\centerline{ {\hfil\hfil
\psfig{figure=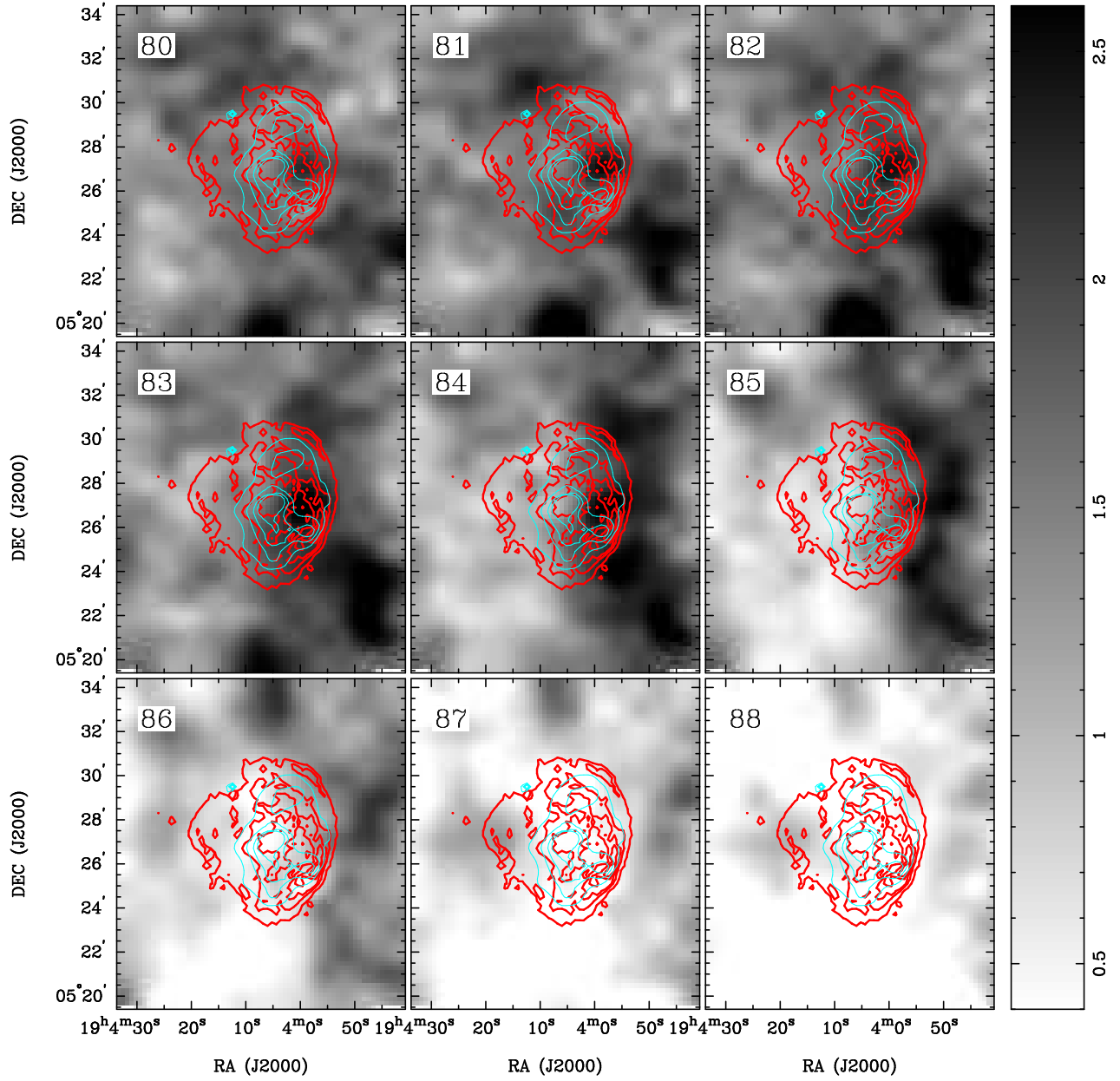,height=6.5in,angle=270, clip=} \hfil\hfil}}
\caption{SRAO \twCO\ ($J$=1--0) intensity channel maps between 80
and 88 km~s$^{-1}$ (smoothed to a resolution of $0'.2$ by
interpolation), overlaid with the 1.0--7.0~keV X-ray ({\em cyan,
thin}) contours and the  1.4~GHz radio ({\em red, thick})
continuum emission contours at the same levels as in
Figure~\ref{f:srao6472}.} \label{f:srao8088}
\end{figure*}

The large field of view ($14'\times15'$) \twCO\ channel maps are
made with the SRAO observation and shown in Figures
\ref{f:srao6472} and \ref{f:srao8088}, overlaid with radio (red
thick) and X-ray (cyan thin) contours. The velocity components
around 69~km~s$^{-1}$ (in 64--72~km~s$^{-1}$ interval) and
around 84~km~s$^{-1}$ (in 80--88~km~s$^{-1}$ interval) both
seem to have positional correlations with SNR 3C~396. The
69~km~s$^{-1}$ MCs are mainly situated in the north of the remnant
[with the 64--67~km~s$^{-1}$ molecular gas in the northeast and
the 68--71~km~s$^{-1}$ gas in the northwest (NW)].
The SNR seems to be
projected roughly within a cavity-like structure of the \twCO\
69~km~s$^{-1}$ component (Figure~\ref{f:srao6472}).
However, the morphology of the 69~km~s$^{-1}$ component appears to
not well match the radio boundary of the remnant; especially,
the 68--71~km~s$^{-1}$ gas appears to overlap the northwestern corner
of the remnant. On the other hand, at around 84~km~s$^{-1}$
(Figure~\ref{f:srao8088}),
a thick wall structure of molecular gas
is seen in the west; the eastern face of the wall at the
85--87~km~s$^{-1}$ interval perfectly follows the western radio
boundary of the remnant.
The radio contours of the remnant seem to coincide with a cavity
at 85--$87\km\ps$.
The CO emission around 84~km~s$^{-1}$ and the remnant's radio
emission both seem to fade out gradually from west to east.

\begin{figure}[tbh!]
\centerline{ {\hfil\hfil
\psfig{figure=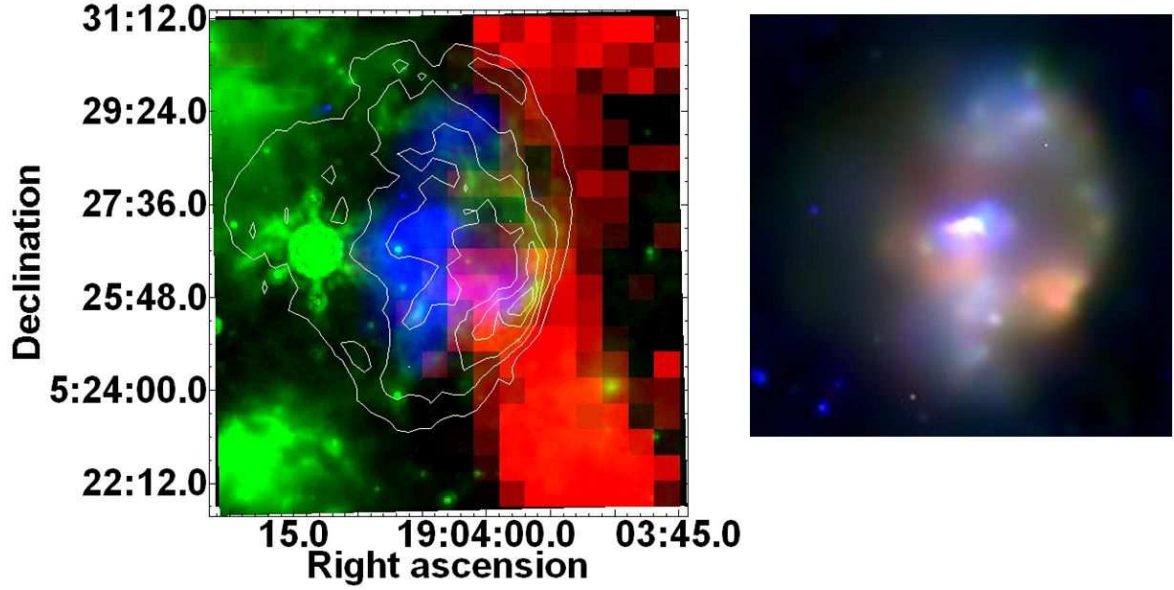,height=3.1in,angle=0, clip=} \hfil\hfil}}
\caption{Left: multiband image of SNR 3C~396. The PMOD $^{12}$CO
intensity map ($>2.2$~K~km~s$^{-1}$ or $11\sigma$) in
85--88~km~s$^{-1}$ interval is in {\em red}, the {\sl Spitzer} 24~$\mu$m
mid-IR emission in {\em green}, and the {\sl Chandra} 1.0--7.0~keV
X-rays (adaptively smoothed with the CIAO program {\em csmooth} to
achieve a S/N ratio of 3) in {\em blue}. Each of the intensity maps
is in logarithmic scale. The five levels of the VLA 1.4~GHz contours
are linear between the 78.6\% and 12.4\% maximum brightness.
Right: {\sl Chandra} ACIS X-ray tri-color
image of SNR 3C~396. The $Si$ (1.65--2.1~keV) is in {\em red}, the
$S$ (2.25--2.6~keV) in {\em green}, and the $hard$ $continuum$
(3.0--7.0~keV) in {\em blue}.} \label{f:otherrgb}
\end{figure}

These characteristics are seen more clearly in the close-up image
(Figure~\ref{f:otherrgb}, left panel) made from the PMOD
observation with a higher resolution. As also shown in
Figure~\ref{f:otherrgb} (left panel), the X-ray/mid-IR/radio
emission in the west of the remnant appears to be well confined by
the western molecular wall around 84~km~s$^{-1}$. Moreover, a
prominent finger/pillar-like MC is seen in the
southwest (SW), with one end intruding inside the SNR border. The
radio and IR brightness peaks of the remnant are located at the
short southwestern section of shell that ``cuts'' the molecular
pillar. Also in the southwestern edge, an X-ray enhancement is
coincident with the radio peak; in Section~\ref{xray} below, we will
show that this X-ray enhancement is exceptionally soft and of
normal metal abundance.

\begin{figure*}[tbh!]
\centerline{ {\hfil\hfil
\psfig{figure=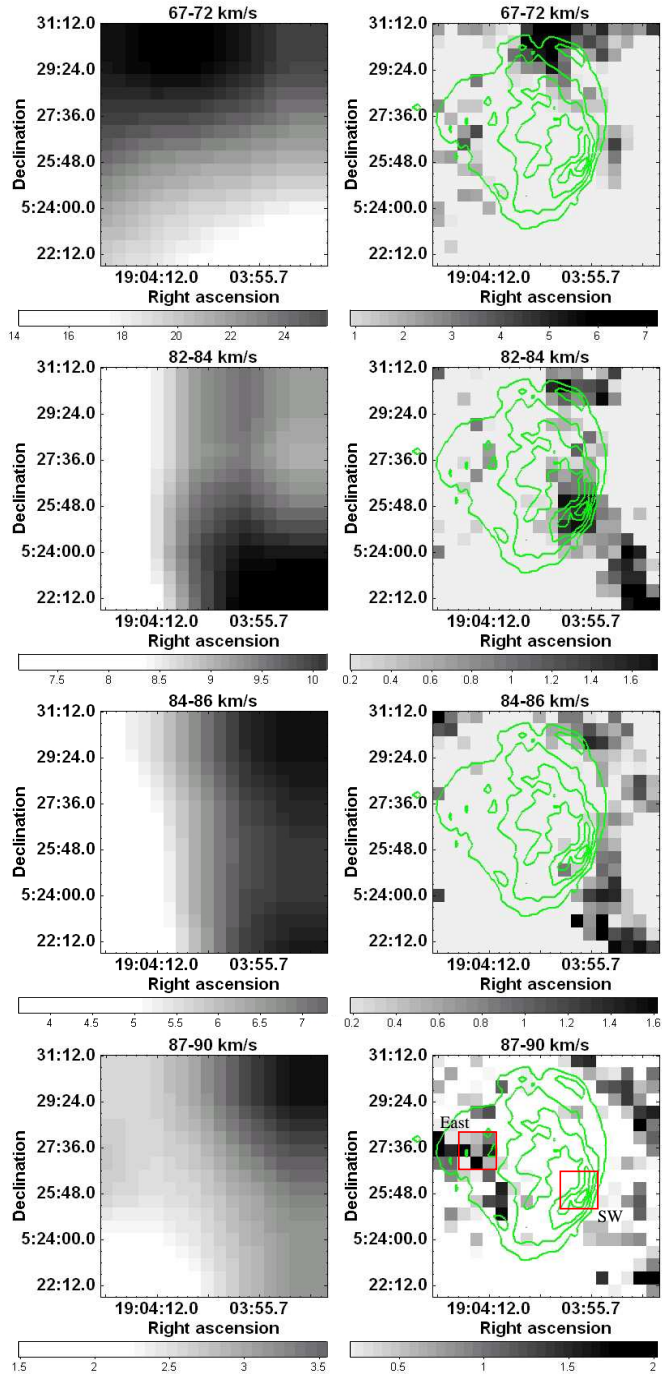,height=7.3in,angle=0, clip=} \hfil\hfil}}
\caption{The four left panels: the PMOD $^{12}$CO
intensity maps in the intervals of 67--72, 82--84, 84--86, and
87--90~km~s$^{-1}$, smoothed to $2'$ to illustrate the large-scale
confusion in this direction. The four right panels: the PMOD $^{12}$CO
maps in the same velocity intervals after unsharp
masking (see text in Section~\ref{HICO}), overlaid by VLA 1.4~GHz
contours at the same levels as in Figure~\ref{f:otherrgb}.
} \label{f:4f}
\end{figure*}

As an attempt to remove the confusion by large-scale density gradient,
we used the unsharp masking method as described in Landecker et
al.\ (1999), in which we subtracted the smoothed \twCO\ intensity
map from the original \twCO\ map. Similar images can also be
produced by subtracting the smoothed and scaled \thCO\ intensity
image from original \twCO\ image. In the ``cleaned" image of the
69~km~s$^{-1}$ MCs, the cavity-like structure is seen again
(Figure~\ref{f:4f}, the right panel in the first line),
but it clearly does not well match the radio boundary of
the remnant.
For the 84~km~s$^{-1}$ MCs, on the other hand, the southwestern
finger/pillar-like intrusion and the western molecular wall can be
seen clearer in the ``cleaned" image (Figure~\ref{f:4f}, the right panel in the second line)
than in the intensity maps. The western boundary of
the SNR appears to perfectly follow the left side of a molecular wall again
(Figure~\ref{f:4f}, the right panel in the third line). A patch of strong
\twCO\ emission around 84~km~s$^{-1}$ is seen in the
NW, where the radio emission of the SNR is slightly
enhanced; this patch can also be discerned in the channel map,
Figure~\ref{f:srao8088}, at $84\km\ps$.
In addition, there is another patch of \twCO\ emission
in 87--90~km~s$^{-1}$ interval in the eastern ``blow-out'' region
of the remnant (Figure~\ref{f:4f}, the right panel in the fourth line); it can also
be discerned in Figure~\ref{f:srao8088} at 87--88~km~s$^{-1}$.

\begin{figure*}[tbh!]
\centerline{ {\hfil\hfil
\psfig{figure=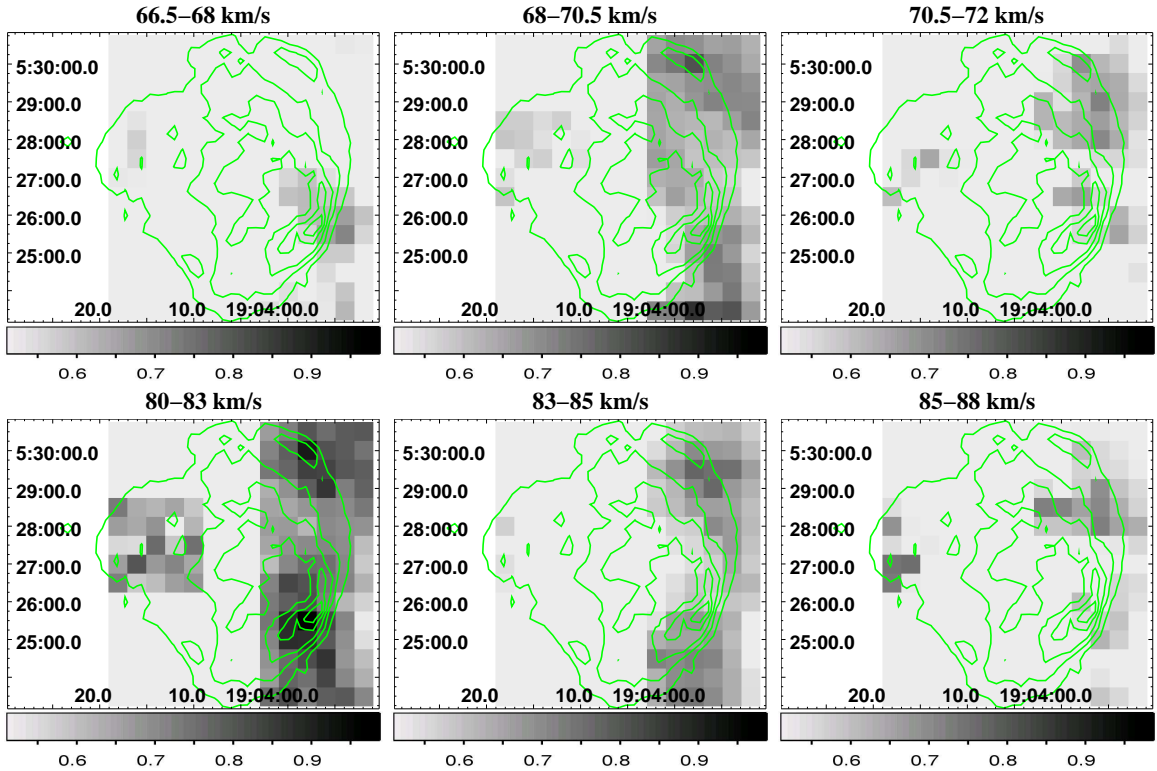,height=4.0in,angle=0, clip=} \hfil\hfil}}
\caption{Maps of \RCO\ intensity ratio around the SNR 3C~396 at
velocity ranges 66.5--68, 68--70.5, 70.5--72, 80--83, 83--85, and
85--88~km~s$^{-1}$ after convolving the SRAO \twCO\ ($J$=2--1)
data to the same beam size as the PMOD \twCO\ ($J$=1--0) data.
Contours show the VLA 1.4~GHz emission at the same levels as in
Figure~\ref{f:otherrgb}.} \label{f:ratio}
\end{figure*}

The integrated CO-line intensity ratio between the different
transitions is suggested as a probe of the SNR--MC interaction
(Seta et al.\ 1998; Jiang et al.\ 2010). The \RCO\ ratio maps
towards 3C396 at 66.5--68, 68--70.5, 70.5--72, 80--83, 83--85,
and 85--$88\km\ps$ (which covers the centers and wings of the
69 and $84\km\ps$ lines, respectively) are presented in
Figure~\ref{f:ratio}.
For the $\sim84\km\ps$ component, the ratios are clearly elevated
to as high as $>0.9$ in 80--$83\km\ps$ interval (compared with
$\sim0.5$--0.7 in average) in the southwestern molecular intrusion
region and the northwestern molecular patch (corresponding to
the two radio brightness peaks, respectively) described above.
For the $\sim69\km\ps$ (66.5--$72\km\ps$) component, although the
ratios are weakly elevated in some boundary regions, the
ratios are all below 0.9.
Therefore the elevated \RCO\ ratios seem to better imply an interaction
of the SNR with the $\sim84\km\ps$ MCs than with the $\sim69\km\ps$
ones.

We have carefully examined the CO spectra
in the field of view and presented the \twCO($J=1$--0), \twCO($J=2$--1),
and \thCO($J=1$--0) spectra extracted from two regions
(in the east and SW, as marked in the right panel in the fourth line of Figure~\ref{f:4f})
in Figure~\ref{f:2spec}.
Around $69\km\ps$, the \thCO\ line profiles are very similar to
the \twCO\ profiles for both the ``East" and ``SW" regions.
Both the \twCO\ and \thCO\ spectra of the ``SW" region have
peaks at 84~km~s$^{-1}$, with broad blue (left) wings in the line
profiles.
Yet, it is difficult to judge whether the blue wings of the 84~km~s$^{-1}$
\twCO\ lines are dynamically broadened or not, due to the significant
blue wing of the \thCO\ line and especially a distinct \thCO\ peak
at $81\km\ps$. \thCO\ emission is usually optically thin and
indicative of quiescent, intrinsically high-column-density MCs.
For the ``East" region, both the \twCO\ and \thCO\ spectra
have peaks at 81~km~s$^{-1}$; notably, there is
an additional broadening in the red (right) wing
(86--90~km~s$^{-1}$) of the 81~km~s$^{-1}$ \twCO\ line, as compared
with little \thCO\ emission in the same velocity interval.
It most likely indicates that the \twCO\ emission in this interval
results from the perturbation suffered by
the 81~km~s$^{-1}$ molecular gas in the east of the remnant.
Although there is a velocity discrepancy between $81\km\ps$ for the
eastern region and $84\km\ps$ for the western region, the undisturbed
clouds in the SNR field may have a wide velocity distribution, e.g.,
the LSR velocity of MCs with which SNR IC~443 interacts is in the range
of $-7$ to $-2\km\ps$ (van Dishoeck et al.\ 1993; Hewitt et al.\ 2006).
Thus the $81\km\ps$ cloud in the east is very likely, together with
the $84\km\ps$ clouds, among the MCs shocked by SNR 3C~396.

The above analysis show that both of the 69 and 84~km~s$^{-1}$
molecular components appear to have spatial correlation with SNR
3C~396, but the latter seems to have much better evidence than the former. We will
further show that the 69~km~s$^{-1}$ clouds are in the foreground of
SNR~3C~396, using the VGPS radio and GRS \thCO\ data.

The 1420 MHz continuum image (Figure~\ref{f:HI}, upper left panel)
displays semicircularly brightened structure in the side roughly
close to the Galactic plane and faint emission in another side.
Following the method used in Tian et al.\ (2007), we produced the
HI spectrum (Figure~\ref{f:HI}, upper right panel) of the radio
peak region of the remnant. The maximum radial velocity of
absorption features in the direction towards 3C~396 is up to the
tangent point LSR velocity (i.e., $\sim$~84~km~s$^{-1}$), and
virtually no significant absorption feature appears at negative
velocities. It indicates that the distance of SNR 3C~396 is
between 6.2~kpc (the tangent point) and 12.5~kpc (the Solar
circle). Here we have used the rotation curve in Clemens (1985)
with $R_0=8.0$~kpc (Reid 1993) and $V_0=220$~km~s$^{-1}$ and
adopted the spiral model in Taylor \& Cordes (1993; Figure~1
therein).

The HI self-absorption (SA) measurement can
provide another clue to the MCs' distances. A single radial
velocity indicates two distances (the near and the far kinematic
distances) in the inner Galaxy along the line-of-sight (LOS). MCs
in the near distance often display HI SA features and many work
have been carried out based on this property
(e.g., Liszt et al.\ 1981; Jackson et al.\
2002; Goldsmith \& Li 2005; Busfield et al.\ 2006; Anderson \&
Bania 2009; Roman-Duval et al.\ 2009).
The HI SA property also seems effective in resolving kinematic
distance ambiguity toward SNR--MC associations, such as 3C~397
(Jiang et al.\ 2010) and CTB~109 (Tian et al.\ 2010).
Figure~\ref{f:HI} (lower panels) shows the VGPS HI and GRS \thCO\
spectra of two regions (as defined in the upper left panel) near
the remnant. The HI SA feature is clearly seen at 69~km~s$^{-1}$
associated with the \thCO\ peak, which indicates that the
69~km~s$^{-1}$ molecular component is most probably located at the
near distance along the LOS, i.e., 4.3~kpc. This component seems
to be a part of the 69.65~km~s$^{-1}$ MC complex
GRSMC~G039.34$-$00.26 at the near side (Roman-Duval et al.\ 2009),
which is apparently outside the distance range 6.2--12.5~kpc
inferred for SNR~3C~396.

Therefore we suggest that SNR~3C~396 is more probably associated
with the MCs at $\sim84$~km~s$^{-1}$. The derived column density
and gas mass of some molecular features (the northwestern patch,
southwestern pillar and the tip, and the western wall
(Figure~\ref{f:coregion})) are presented in Table~\ref{l:CO}. In
the derivation of these parameters, two methods are used and
similar results are yielded. In the first method, the H$_2$ column
density is calculated by adopting the mean CO-to-H$_2$ mass
conversion factor $N({\rm H}_2)/W(^{12}$CO)$
\approx1.8\times10^{20}$~cm$^{-2}$K$^{-1}$km$^{-1}$s (Dame et al.\
2001). In the second method, on the assumption of local
thermodynamic equilibrium and the excitation temperature
$\sim10$~K, the \thCO\ column density is converted to the H$_2$
column density using the relation $N($H$_2)$=$7\times10^5N$(\thCO)
(Frerking et al. 1982). Assuming a cuboidal volume for the wall
region (Figure~\ref{f:coregion}) with size
$4'.5\times4'.5\times9'.5$,
the mean molecular density of the western wall is inferred as
$\sim170\du^{-1}~{\rm H}_{2}$~cm$^{-3}$.

\subsection{X-ray Properties of 3C~396}\label{xray}

Next, we investigate the positional relation of SNR~3C~396 X-ray
emission with the CO intensity distribution in
detail. In general, we extracted {\sl Chandra} X-ray spectra from
five X-ray bright regions (as defined in
Figure~\ref{f:xrayregion}) using the CIAO3.4 script {\sl
specextract}. The background spectrum was extracted from four
nearby regions (dashed circles in
Figure~\ref{f:xrayregion}). The five on-source spectra were
adaptively regrouped to achieve a background-subtracted
signal-to-noise ratio (S/N) of 5 per bin. Then we used a nonequilibrium
ionization (NEI) model to fit the spectra of the remnant.
The XSPEC spectral fitting package was used throughout. For the foreground
absorption, the cross sections from Morrison \& McCammon (1983) were used.
The parameters of the fitting are listed in
Table~\ref{l:xfit} with the 90\% confidence ranges.

The five spectra all contain distinct lines of Si XIII
($\sim1.85\keV$) and S XV ($\sim2.45\keV$), but Ar XVII
($\sim3.12\keV$) and Ca XIX ($\sim3.86\keV$) lines are prominent
only in the northern and southern regions
(Figure~\ref{f:xrayspec}). The model of our spectral fit generates
a variable hydrogen column density from $4.0\E{22}$~cm$^{-2}$ to
$6.5\E{22}$~cm$^{-2}$. These values are roughly consistent with
$4.65\E{22}$~cm$^{-2}$ from the diffuse X-ray spectral fitting
with {\sl ASCA} X-ray observation (Harrus \& Slane 1999) and
$5.3\E{22}$~cm$^{-2}$ from the entire central nebula spectral
fitting with {\sl Chandra} X-ray observation (Olbert et al.\
2003). The northern and southern regions have relatively high
temperatures ($\ga1\keV$) and slight overabundance of Si, S, and
Ca. The temperatures of the other three regions are relatively
low, around $0.7\keV$. The western edge is also S enriched and,
notably, only the southwestern region is found to be of normal
abundance.
The ionization timescales, $n_et$, in the northern, southern,
and western regions are less than $5.5\E{11}$~s~cm$^{-3}$ but
seems to be higher in the southwestern region
($>4\E{11}$~s~cm$^{-3}$).

An estimate of the hot-gas hydrogen-atom density $\nX$ of these
regions is made using the volume emission measure, $fn_{e}\nX V$,
derived from the normalization of the XSPEC model, where
$n_e$ is the electron density, $V$ the volume of the emitting
region, and $f$ the volume filling factor of the hot-gas
($0<f<1$). Here we have assumed $n_{e}\approx1.2\nX$ and
oblate spheroids for the defined elliptical regions
(Figure~\ref{f:xrayregion}). The derived physical parameters are
listed in Table~\ref{l:xray}. If we adopt
$\nX=2f^{-1/2}\du^{-1/2}\cm^{-3}$ as the mean density of SNR's
hot-gas (where $\du=d/6.2\kpc$ is used for scaling;
Section~\ref{distance}) and assume an oblate spheroid
($2'.7\times3'.9\times3'.9$) for the entire diffuse thermal X-ray
emitting volume, $V=3.0\times10^{58}~f\du^{3}$~cm$^{3}$, the mass
of the X-ray emitting hot-gas of SNR 3C~396 is
$\sim70f^{1/2}\du^{5/2}\Msun$.

In the {\sl Chandra} ACIS three-color image
(Figure~\ref{f:otherrgb}, right panel), the X-ray emission is hard
in the central PWN region and the northern and southern regions
(dominated by the {\em blue} color), relatively soft along the
western boundary (dominated by the {\em green} color), and softest
in the southwestern edge (seen as an {\em orange} patch). This is
consistent with the distribution of hot-gas temperatures derived
above. The softest patch is coincident with the tip of the
$\VLSR\sim84\km\ps$ molecular finger/pillar.

\section{DISCUSSION}
\subsection{Association of SNR 3C~396 with the $\VLSR
 \sim$84~km~s$^{-1}$ MCs and the Distance}\label{distance}

The cavity-like \twCO\ structure at $\VLSR\sim69\km\ps$ seems
consistent with the GRS \thCO\ data (Lee et al.\ 2009; Figure~7
therein). Based on such a structure at $\VLSR\sim69\km\ps$ that seems
to roughly surround SNR 3C~396, Lee et al.\ (2009) suggest an
association of 3C~396 with the $\sim69\km\ps$ MCs and thus adopt
a distance 8.5~kpc.
However, the HI SA property of the 69~km~s$^{-1}$ molecular
component indicates that it is most likely to be in the foreground
($\sim4.3$~kpc) toward SNR 3C~396,
while the 21~cm HI absorption feature suggests the SNR being
located in the distance range 6.2--12.5~kpc (Section~\ref{HICO}).
It thus appears difficult to associate the $\sim69\km\ps$ MCs with the
remnant. In our \twCO\ observation, we indeed find that the
cavity-like structure of the 69~km~s$^{-1}$ molecular component does
not well match the morphology of the remnant (Section~\ref{HICO}).

Besides the 69~km~s$^{-1}$ MCs, however, we have found that
the western boundary of the SNR is perfectly confined by the
western molecular wall at $\VLSR\sim84\km\ps$ and the
multiwavelength (X-ray, mid-IR, and radio) properties are consistent
with the presence of the $84\km\ps$ MCs (Section~\ref{HICO}).

\begin{figure*}[tbh!]
\centerline{ {\hfil\hfil
\psfig{figure=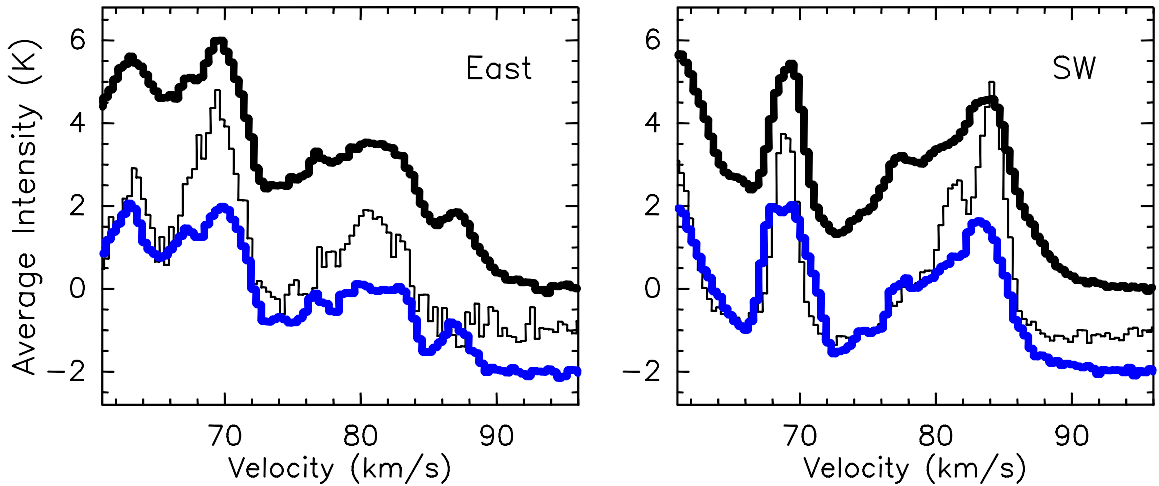,height=2.5in,angle=270, clip=} \hfil\hfil}}
\caption{The CO spectra in 61--96~km~s$^{-1}$ interval for the
eastern [the ``East" region in Figure~\ref{f:4f}, 2.25~arcmin$^2$
centered at $(\RA{19}{04}{14},\Dec{05}{27}{30})$] and the
southwestern [the ``SW" region in Figure~\ref{f:4f},
2.25~arcmin$^2$ centered at $(\RA{19}{03}{58},\Dec{05}{25}{30})$]
regions of SNR 3C~396. The thick {\em black}, thick {\em blue},
and thin lines are for the PMOD \twCO\ ($J$=1--0), SRAO \twCO\
($J$=2--1), and PMOD \thCO\ ($J$=1--0) (multiplied by a factor of
4) emission, respectively. The profiles of \thCO\ ($J$=1--0) and
\twCO\ ($J$=2--1) are shifted by $-1$ and $-2$~K, respectively.
The \twCO\ ($J$=2--1) spectra are not convolved from beam size
$48''$ to the \twCO\ ($J=1$--0) beam size $54''$.} \label{f:2spec}
\end{figure*}

\begin{figure*}[tbh!]
\centerline{ {\hfil\hfil
\psfig{figure=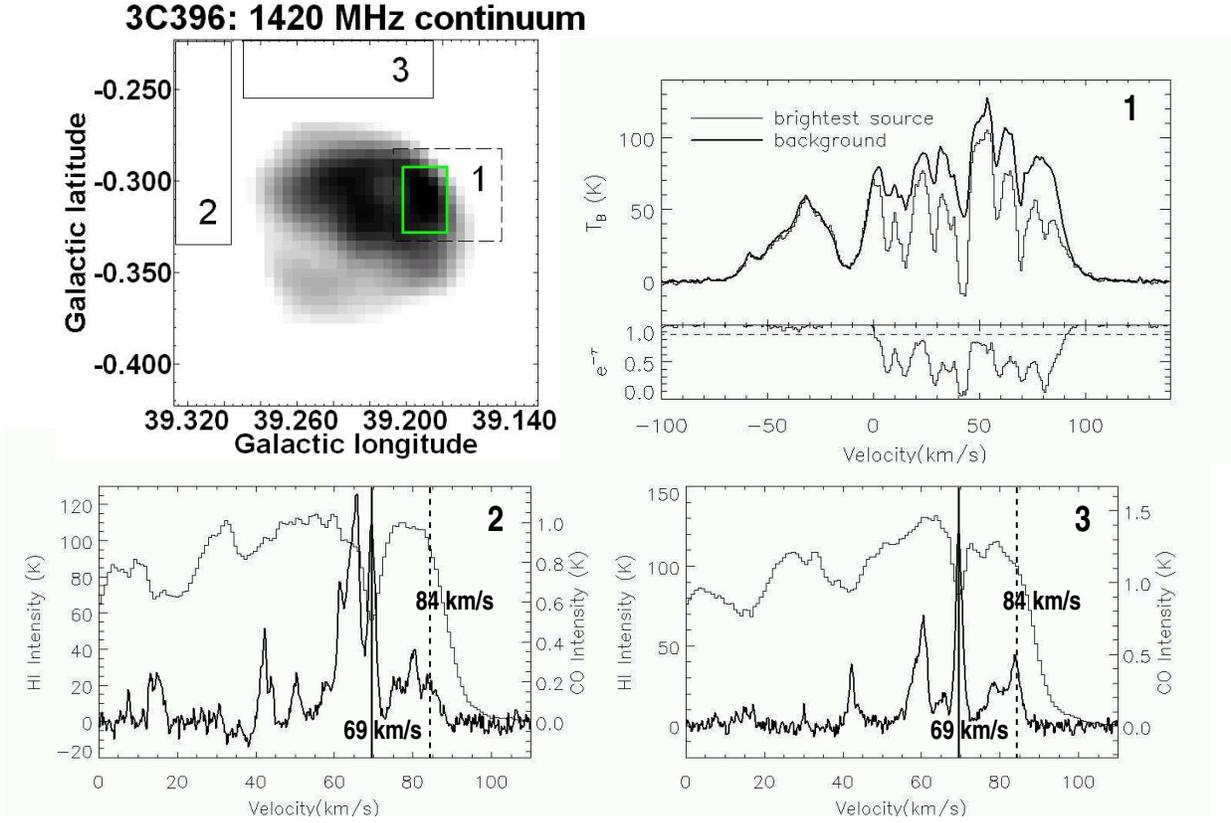,height=4.3in,angle=0, clip=} \hfil\hfil}}
\caption{Upper left panel: the VLA 1.4~GHz continuum image of SNR
3C~396. Upper right panel: the HI spectra extracted from box 1.
The HI spectrum of the bright source is from the solid-line box
area, and the background spectrum is from the dashed-line box
excluding the solid box area. The dashed line indicates
5~$\sigma$ derivation for velocities $-100\leqslant$ $V_{\rm LSR}\leqslant$
$-$50~km~s$^{-1}$ and 110$\leqslant$ $V_{\rm LSR}\leqslant$
140~km~s$^{-1}$.
Lower panels: the VGPS HI (thin) and GRS $^{13}$CO (thick) spectra
extracted from boxes 2 and 3 (see the upper left panel), respectively.
The vertical solid and dashed lines indicate the position of 69 and
84~km~s$^{-1}$, respectively.
} \label{f:HI}
\end{figure*}

\begin{figure}[tbh!]
\centerline{ {\hfil\hfil
\psfig{figure=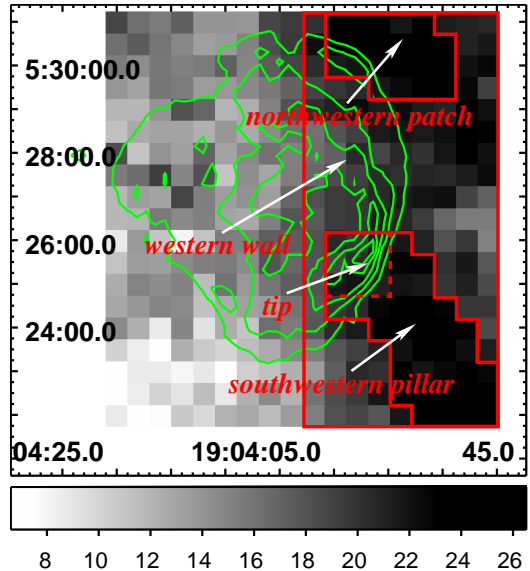,height=3.0in,angle=0, clip=} \hfil\hfil}}
\caption{PMOD \twCO\ intensity map in
82--90~km~s$^{-1}$ interval, labelled with the defined regions for the
parameter derivation of the 84~km~s$^{-1}$ molecular gas.
Contours show the 1.4~GHz radio continuum emission towards the
remnant at the same levels as in Figure~\ref{f:otherrgb}.}
\label{f:coregion}
\end{figure}

\begin{figure}[tbh!]
\centerline{ {\hfil\hfil
\psfig{figure=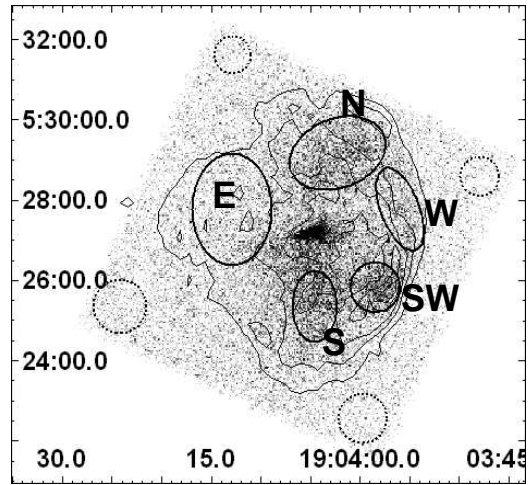,height=2.5in,angle=0, clip=} \hfil\hfil}}
\caption{{\sl Chandra} X-ray raw image of SNR~3C~396, labelled
with the solid (dotted) defined regions for the source
(background) X-ray spectrum extraction. The point sources (except
the central pulsar) detected with a wavelet source-detection
algorithm have been removed. The contours are the VLA 1.4~GHz
radio continuum emission at the same levels as in
Figure~\ref{f:otherrgb}.} \label{f:xrayregion}
\end{figure}

\begin{figure*}[tbh!]
\centerline{ {\hfil\hfil
\psfig{figure=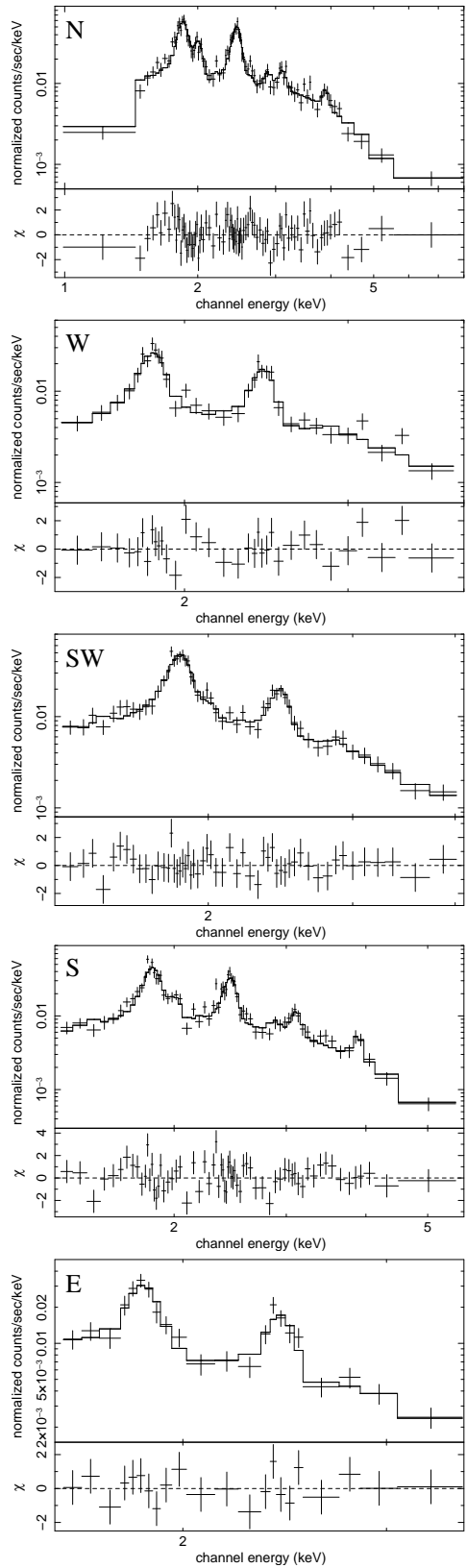,height=8.5in,angle=0, clip=} \hfil\hfil}}
\caption{{\sl Chandra} ACIS X-ray spectra of the SNR~3C~396 (taken
from a 93.7~ks data), fitted with the {\em vnei} model.}
\label{f:xrayspec}
\end{figure*}

The eastern \twCO\ patch (seen in Figure~\ref{f:srao8088}, at
87--$88\km\ps$; Figure~\ref{f:4f}, the right panel in the fourth line) is shown to
have a redshifted broadening (87--90~km~s$^{-1}$) in the \twCO\
spectral line profiles (Figure~\ref{f:2spec},
left panel) and thus probably represents the molecular gas
perturbed by the SNR.
The molecular line broadening has widely been used as a kinematic
evidence of interaction between SNR shocks and MCs (see Jiang et al.\
2010), such as in the studies of prototype interacting SNRs IC~443
(DeNoyer 1979), W28 (Arikawa et al.\ 1999), HB21 (Byun et al.\ 2006),
and G347.3-0.5 (Moriguchi et al.\ 2005).
More recently, broadened line profiles of optically thin \twCO\ lines were
also successively found in our serial work on SNR--MC interaction
(Kes~69, Zhou et al.\ 2009; Kes~75, Su et al.\ 2009; and 3C~397,
Jiang et al.\ 2010).

Although the broad blue wings of the $\sim84\km\ps$ \twCO\ lines
of the western region of 3C~396 cannot be exclusively determined to be
dynamically broadened profiles (Section~\ref{HICO}; Figure~\ref{f:2spec}, right panel),
the relatively enhanced \RCO\ ratios in such wings (80--$83\km\ps$)
in the northwestern and southwestern boundaries (Figure~\ref{f:ratio}, lower left panel)
coincide with the southwestern molecular intrusion region and the
northwestern molecular patch, and also with two radio brightness peaks,
respectively, which is
in favor of a perturbation of the western molecular gas by the SNR expansion.
We note that the maximum ratio obtained here is not higher than 1,
while higher values are observed in other interacting SNRs
(e.g., W44, IC~443, HB21, G349.7+0.2, and G18.8+0.3, see Dubner et al.\ 2004
and references therein).
Two main reasons may explain why the ratio value in
80--$83\km\ps$ interval does not exceed unity:
1.~The filling factor of the disturbed \twCO\ ($J$=2--1) emitting region
may be small in the $48''$ beam size in 3C~396 compared to that for
\twCO\ ($J$=1--0), and the line ratios could thus be diluted.
2.~The location of 3C~396 at the tangent point makes the environment
much complicated and may cause unignorable velocity crowding
in the interval of interest, with the line wings difficult
to differentiate. Actually, e.g., there is a distinct cloud represented by
the \thCO\ line peak at $81\km\ps$ in the southwestern molecular intrusion
region (see Figure 6, right panel). Confusion of the \twCO\ emission
of such gas will decrease the \RCO\ ratios in the velocity interval.

In multiwavelength morphologies, both the X-ray and radio emissions are bright
in the western half and faint in the eastern half
(Figures~\ref{f:otherrgb} and \ref{f:xrayregion}), which is probably because of the
large-scale density gradient roughly perpendicular to the
Galactic plane.
The bright side of the remnant is close to the Galactic
plane (see Figure~\ref{f:HI}, upper left panel, in Galactic coordinates).
The bright X-ray, IR and radio emissions in the
west are well confined by the thick molecular wall of the
$\sim84$~km~s$^{-1}$ component, which may suggest that the SNR
expansion is hampered there.
This scenario is strengthened by the radio and IR brightness peaks
and the soft X-ray patch
in the southwestern edge, which are coincident with the ``tip"
of the intruding finger/pillar of molecular gas.
When the SNR shock hits the dense molecular pillar or clumps, it will be
decelerated and thus the magnetic field be compressed and amplified.
This can account for the prominent radio enhancement and sharp edge there.
Actually, the magnetic field in the SW of the remnant is found to
be essentially tangential to the shell,
which is probably because the random interstellar magnetic field
has been chiefly swept up into the accumulated gas and compressed
along with the remnant's shell (Patnaik et al.\ 1990).
[Parenthetically, the tangential field (Patnaik et al.\ 1990)
and the slightly enhanced radio emission in the northwestern edge are
consistent with the presence of the dense molecular clump in the NW
(Section~\ref{HICO}).]
In X-rays, the low electron temperature, normal metal
abundance, and large ionization timescale
(as obtained from our spectral fitting) of the soft southwestern
X-ray patch can be naturally interpreted by the interaction
with the dense molecular pillar.
The IR morphology can also be accounted for with the scenario of
shock interaction with western molecular gas, as discussed in the following.

In IR, there is a spatial separation between the inner near-IR
1.64~$\mu$m [FeII] and outer 2.12~$\mu$m H$_2$ emission filaments along
the western edge of the remnant (Lee et al.\ 2009; Hewitt et al.\ 2009).
Lee et al.\ (2009) interprets this morphology due to
the SNR shock overtaking the wind of a red supergiant (RSG), giving
rise to [FeII] emission and the interaction with a dense
circumstellar bubble, giving rise to \HH\ emission.
Such a RSG wind bubble 5--7~pc in radius invokes a 25--$35\Msun$ progenitor.
This mass assumption is different from our estimate of 13--$15\Msun$
derived from the X-ray emitting metal abundances (see below in Section~\ref{progenitor})
and is not in agreement with the mass range $\la25\Msun$ calculated for
the neutron star progenitors with metallicity $Z\la\Zsun$ (Heger et al.\ 2003).
The 13--$15\Msun$ progenitor of type-IIP supernova (see Section~\ref{progenitor})
can only blow a RSG wind bubble of radius $\la1$~pc (Chevalier 2005).
However, the spatially separated [FeII] and H$_2$ emission can be
interpreted in term of SNR--MC interaction. As Reach et al.\
(2005) and Hewitt et al.\ (2009) suggest,
the H$_2$ emission can arise from the shock interaction with dense
clumps while the [FeII] emission can arise from the
same shock encountering the less dense material at the clumps' surface
or from the ejecta.

The bright 24~$\mu$m mid-IR emission (Figure~\ref{f:otherrgb}) may chiefly come from the dust grains
and even possibly from the ionic and molecular species in the shocked
gas (e.g., Tappe et al.\ 2006; Su \& Chen 2008) when the western
molecular wall is hit by the blast wave.
A spinning dust component is invoked to account for the spectral
excess at 33~GHz (Scaife et al.\ 2007), and it seems
that the needed dust may lie in the MCs and be heated by the shocked
gas.

All these evidences strongly point to
a conclusion that SNR 3C~396 is associated, and in physical contact,
with the $\sim84\km\ps$
MCs. The LSR velocity 84~km~s$^{-1}$ corresponds to a
Galactic location near the tangent point, at a distance of
$\sim6.2$~kpc; this distance is consistent with the range
[6.2--12.5~kpc (Section~\ref{HICO})] inferred from the HI absorption data.
We also notice that a distance of
6.5~kpc suggested by Hewitt et al.\ (2009) based on the hydrogen
column is very similar to our result.
Hereafter, we write the distance to 3C~396 as
$d=6.2\du\kpc$ for scaling.

It was queried by Patnaik et al.\ (1990) whether the HII complex
NRAO~591 (or U39.25$-$0.07), which looks projectively near 3C396,
is associated with the SNR.
Since the complex is at $V_{\rm LSR}$= 23~km~s$^{-1}$ with
the far side distance of 12.1~kpc (Anderson \& Bania 2009),
it is not associated with the remnant.
Instead, it is very likely to be associated with the 22.46~km~s$^{-1}$
GRSMC G039.24$-$00.06 at the far side (Roman-Duval et al.\ 2009).

\subsection{SNR physics of 3C~396}\label{physics}
\subsubsection{Global evolution}

According the above multiwavelength analysis, a general scenario
for SNR~3C396 can be established as this:
the remnant collides with a molecular wall in the west, flanking
a southwestern pillar of molecular gas, and expands
in a low density region in the east (but encountering a clump of
backside molecular gas).

The global dynamical evolution of the remnant can be inferred from
the X-ray properties.
Adopting $kT_{\rm X}\sim0.9$~keV as the average postshock temperature
in the interclump medium,
we have the mean blast shock velocity $v_s=[16kT_{\rm X}/(3\bar{\mu}m_{\rm
H})]^{1/2}\sim8.7\E{2}$~km~s$^{-1}$, where $m_{\rm H}$ is the
hydrogen atom mass and $\bar{\mu}=0.61$ is the average atomic
weight.
Assuming an adiabatic expansion and adopting the
SNR radius $r_s\sim3'.9\sim7.0\du$~pc and the mean density of the
interclump medium $n_0\sim\nX/4\sim1f^{-1/2}\du^{-1/2}$cm$^{-3}$,
the SNR's age is estimated
as $t=(2r_s)/(5v_s)\sim
3\du$~kyr and the explosion energy is
$E=(25/4\xi)(1.4n_0 m_{\rm
H})v_s^2r_s^3\sim6(n_0/1\cm^{-3})\E{50}\du^{5/2}$~erg (where
$\xi=2.026$).
The dynamical age estimated here is consistent to the
ionization ages (the time since passage of the shock wave) of the
X-ray emitting gas obtained from the spectral fit (Table~\ref{l:xray}),
except the region ``W" for which the ionization age seems relatively
low.

\subsubsection{Pressure discrepancy}

The thermal pressure of the X-ray emitting gas is about
$p_{\rm h}\sim1\times10^{-8}f^{-1/2}\du^{-1/2}$~dynes~cm$^{-2}$ (Table~\ref{l:xray}),
which is about two order of magnitudes lower than that of the shock
ram pressure $p_{\rm cl}\sim3\times10^{-6}$~dynes~cm$^{-2}$,
as derived from the H$_2$ emission of the {\sl Spitzer} IRS
observation towards the southwestern region (Hewitt et al.\ 2009).
Similar pressure discrepancies have also been
found in the cases of IC~443 (Moorhouse et al.\ 1991) and
3C~391 (Reach \& Rho 1999). Such discrepancies can be explained
if the western molecular gas is clumpy and there is a radiative
shell between the hot gas and the dense
molecular clumps (Chevalier 1999).
A radiative shock has been revealed by the {\sl Spitzer} IRS slit observation
(Hewitt et al.\ 2009).
On the assumption of a crude pressure balance between the hot gas
and the radiative shell, we have (Chevalier 1999)
\[\frac{p_{\rm cl}}{p_{\rm h}}\sim\frac{n_{\rm cl}}{n_0}
\left[1+\left(\frac{n_{\rm cl}}{n_{\rm rs}}\right)^{1/2}\right]^{-2},\]
where $n_{\rm cl}$ is the hydrogen atom density in the clump,
$n_0$ the interclump density,
and $n_{\rm rs}$ the density in the radiative shell.
The gas density in the dense clump, $n_{\rm cl}\sim1\E{4}\cm^{-3}$,
was estimated for the southwestern region from the H$_2$ lines by two
groups of authors (Lee et al.\ 2009; Hewitt et al.\ 2009).
Considering that the X-ray emission represents the shock interaction
with the interclump gas, we may assume $n_0$ is about 1/4 the density
of the hot gas ($\nX$, see Table~\ref{l:xray}) and thus is of order
$1\cm^{-3}$ for the southwestern region.
Hence we obtain $n_{\rm rs}\sim 4\E{2}\cm^{-3}$.
This number is similar to the electronic density ($\sim3\E{2}\cm^{-3}$)
inferred from ionic lines (Hewitt et al.\ 2009).

\subsubsection{Supernova ejecta and progenitor} \label{progenitor}

\begin{figure}[tbh!]
\centerline{ {\hfil\hfil
\psfig{figure=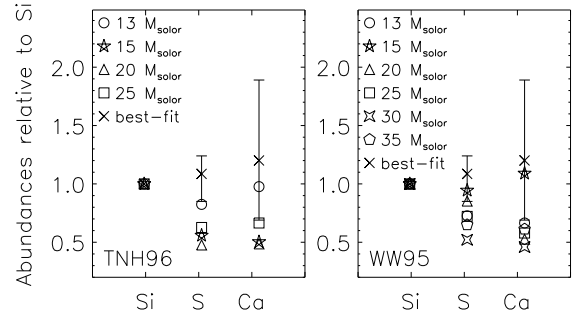,height=2.0in,angle=0, clip=} \hfil\hfil}}
\caption{The best-fit abundances of the X-ray emitting gas in the
north and south (the ``N+S" column in Table~\ref{l:xfit}) relative
to Si, compared with the predictions from core-collapsed SN models
(TNH96 and WW95). } \label{f:mass}
\end{figure}

In the north and south, the $\sim84\km\ps$ CO emission is very weak,
the SNR does not seem to encounter the molecular wall,
and the X-ray emitting gas is found to be metal enriched (Table~\ref{l:xfit}).
We assume that the observed pattern of metal enrichment can
reflect the composition of the supernova ejecta and thus be used to
constrain the mass of the progenitor star (following, e.g., Rakowski et al.\
2001; Chen et al.\ 2006). As the presence of a
pulsar implies a core-collapse nature of the SN, we plot the
relative metal abundances predicted from such SN models for stars
with normal metallicities ($Z=\Zsun$) (Thielemann, Nomoto, \& Hashimoto 1996,
hereafter TNH96; Woosley \& Weaver 1995, hereafter WW95) in Figure~\ref{f:mass}.
The enrichment appears compatible with the WW95 model for a progenitor mass
of $M=15\Msun$ or the TNH96 model for a progenitor
mass of $M=13\Msun$. A 13--$15\Msun$ progenitor is a B1--B2 star (Snow 1982)
and ends its life as a type-IIP supernova (Heger et al.\ 2003).

\section{CONCLUSIONS}

We have investigated the molecular gas environment of the semicircular
young composite SNR 3C~396 in multiwavelengths,
principally with \twCO\ ($J$=1--0 and $J$=2--1) and \thCO\ ($J$=1--0)
observations in SRAO and PMOD, and
performed a {\sl Chandra} spatially resolved thermal X-ray spectroscopic
study of the SNR.
The main conclusions are summarized as follows.
\begin{enumerate}

\item We confirmed a cavity-like structure of molecular gas at
$\sim67$--$72\km\ps$ in the 3C~396 SNR region, which was noticed by Lee
et al.\ (2009) based on the GRS \thCO\ ($J$=1--0) data.  It is
most probably to be located in the near distance ($\sim4.3$~kpc)
according to the HI SA measurement.

\item We find a series of evidence suggesting that the molecular gas at
$\sim$~84~km~s$^{-1}$ is in physical contact with SNR 3C~396, which
can reasonably explain the multiwavelength properties of the remnant.
Around this LSR velocity, a molecular wall appears to perfectly
confine the western boundary of the SNR, which is bright
in X-ray, IR, and radio emission. The CO emission
fades out from west to east, indicating
that the eastern region is of low gas density, accounting for the
radio ``blow-out'' morphology in the east of the remnant.
A molecular pillar is revealed in the SW, with one end
intruding inside the SNR border. The shock interaction with this
``pillar tip" can explain the X-ray and radio enhancement in the
SW and some infrared filaments there. The SNR--MC interaction is
also favored by the relatively elevated \twCO\ $J$=2--1/$J$=1--0 line ratios
at 80--$83\km\ps$ in the southwestern ``tip" and the molecular patch on the
northwestern boundary.
Redshifted broadening (86--$90\km\ps$) is detected in the \twCO\
spectral line profiles of the eastern $81\km\ps$ molecular patch and may be
the kinematic evidence for shock--MC interaction.

\item The establishment of the association between 3C~396 and the
84~km~s$^{-1}$ MCs results in a determination of the kinematic
distance at $\sim6.2\kpc$ to the remnant, which agrees with a
location at the tangent point along the LOS.

\item The diffuse thermal X-ray emitting gas is found to be metal
enriched except in the southwestern X-ray enhancement, and the
metal enrichment in the north and south implies that the supernova
progenitor is a 13--$15\Msun$ B1--B2 star.

\item The X-ray spectral analysis suggests an age of the remnant
of $\sim3$~kyr. There is a pressure discrepacy between the X-ray
emitting gas and the IR emitting cloud shock in the west,
which can be explained with the inhomogeneous molecular gas
and radiative shells between the hot gas and the dense molecular
clumps.
\end{enumerate}

\begin{acknowledgements}
The authors are thankful to the staff members of the SRAO and the
Qinghai Radio Observing Station at Delingha for their support in
observation and to Lawrence Rudnick for providing the VLA data of
SNR 3C~396. We acknowledge the use of the VGPS and GRS data; the
National Radio Astronomy Observatory is a facility of the National
Science Foundation operated under cooperative agreement by
Associated Universities, Inc. This work is supported by the NSFC grants
10621303, 10673003, and 10725312 and the 973 Program grant 2009CB824800.
\end{acknowledgements}

\begin{center}
\begin{deluxetable}{ccc}
\tablecaption{Parameters of the $\sim84\km\ps$ Molecular Gas }
\tablehead{\colhead{Region\tablenotemark{a}}
          &\colhead{$N({\rm H}_2)$(10$^{21}$cm$^{-2})$\tablenotemark{b}}
          &\colhead{$M(\Msun)$\tablenotemark{b}}}
\startdata Northwestern clump & $4.4/4.1$ & $1.9\times10^3\du^{2}/1.8\times10^3\du^{2}$ \\
    Pillar & $4.6/4.5$ & $3.7\times10^3\du^{2}/3.6\times10^3\du^{2}$ \\
    Pillar tip & $4.3/4.4$ & $6.9\times10^2\du^{2}/7.1\times10^2\du^{2}$ \\
    Western wall & $4.2/3.8$ & $1.3\times10^4\du^{2}/1.2\times10^4\du^{2}$ \\
\enddata
\tablenotetext{a}{Regions are defined in Figure~\ref{f:coregion}.}
\tablenotetext{b}{See text for the two methods used for the column
density derivation.} \label{l:CO}
\end{deluxetable}
\end{center}

\begin{center}
\begin{deluxetable}{ccccccc}
\tablecaption{{\sl Chandra} X-ray Spectral Fits with 90\%
Confidence Ranges} \tablehead{\colhead{Region}
          &\colhead{N}
          &\colhead{W}
          &\colhead{SW}
          &\colhead{S}
          &\colhead{E}
          &\colhead{N+S}}
\startdata $N_{\rm H}$(10$^{22}$cm$^{-2}$) &$4.84^{+0.30}_{-0.13}$
&$5.50^{+0.97}_{-0.92}$ &$4.42^{+0.41}_{-0.39}$
&$4.74^{+0.26}_{-0.33}$ &$4.95^{+1.10}_{-0.83}$
&$4.82^{+0.21}_{-0.20}$ \\
$kT_{\rm X}$(keV) &$1.33^{+0.20}_{-0.19}$
&$0.71^{+0.25}_{-0.15}$ &$0.66^{+0.10}_{-0.08}$
&$1.04^{+0.11}_{-0.14}$ &$0.68^{+0.37}_{-0.19}$
&$1.21^{+0.13}_{-0.13}$ \\
${\rm Si}$ &$1.66^{+0.28}_{-0.27}$  &$0.92^{+0.31}_{-0.23}$
&$1.23^{+0.23}_{-0.24}$ &$1.42^{+0.36}_{-0.27}$ & 1\tablenotemark{a}
&$1.56^{+0.21}_{-0.21}$ \\
${\rm S}$  &$1.73^{+0.29}_{-0.27}$  &$1.99^{+0.77}_{-0.67}$
&$1.20^{+0.32}_{-0.28}$ &$1.60^{+0.40}_{-0.27}$ & 1\tablenotemark{a}
&$1.63^{+0.22}_{-0.21}$ \\
${\rm Ca}$ &$1.91^{+1.16}_{-0.86}$  & 1\tablenotemark{a}
& 1\tablenotemark{a} &$3.08^{+1.93}_{-1.45}$ & 1\tablenotemark{a}
&$1.98^{+0.93}_{-0.86}$ \\
$\tau$(10$^{11}$s~cm$^{-3}$) &$2.19^{+1.50}_{-0.63}$
&$0.42^{+0.46}_{-0.10}$
& $>$4.06 &$2.69^{+2.79}_{-1.09}$ & $>$0.84 & $2.34^{+1.29}_{-0.74}$  \\
$f n_e \nX V/\du^2 (10^{57}$~cm$^{-3})$ &$1.26^{+0.48}_{-0.39}$ &
$2.73^{+4.32}_{-2.73}$ &$2.35^{+1.55}_{-0.88}$
&$1.28^{+0.52}_{-0.38}$ &$2.34^{+0.63}_{-2.34}$
&$2.50^{+0.71}_{-0.52}$ \\
$\chi_r^{2}$(dof) &1.27(78) &1.21(28)
&0.66(48) &1.40(59) &1.14(19) &1.54(101)\\
\enddata
\tablenotetext{a}{Fixed to the solar abundance.}
\label{l:xfit}
\end{deluxetable}
\end{center}

\begin{center}
\begin{deluxetable}{cccccc}
\tablecaption{Parameters of the Hot Gas}
\tablehead{\colhead{Region}
          &\colhead{N}
          &\colhead{W}
          &\colhead{SW}
          &\colhead{S}
          &\colhead{E}}
\startdata ${\rm Volume}
(10^{56}f\du^{3}$~cm$^{3})\tablenotemark{a,b}$ &$10.1$ &$4.4$
&$1.8$ &$3.5$
&$14.5$\\
$\nX(f^{-1/2}\du^{-1/2}$~cm$^{-3}$) &$1.0$ &$2.3$ &$3.3$
&$1.7$ &$1.2$\\
${\rm Mass} (f^{1/2}\du^{5/2}~M_{\odot})$ &$1.2$ &$1.2$
&$0.7$ &$0.7$ &$2.0$ \\
${\rm Ionization~Age} (10^{3}f^{1/2}\du^{1/2}$~yr)  &$5.6$ &$0.5$
&$>3.2$ &$4.0$ &$>1.9$  \\
${\rm Pressure} (10^{-9}f^{-1/2}\du^{-1/2}$~dynes~cm$^{-2})$
&$6.1$ &$7.3$
&$9.7$  &$8.1$ &$3.5$  \\
\enddata
\tablenotetext{a}{Here $f$ is the filling factor of the X-ray
emitting gas and $\du$ is the distance to the remnant in units of
6.2~kpc (Section~\ref{distance}).} \tablenotetext{b}{In the estimate of
the volumes, we assume the oblate spheroids for elliptical regions
N, W, SW, S, and E, with half axes $0'.88\times1'.26\times1'.26$,
$0'.49\times1'.11\times1'.11$, $0'.63\times0'.63\times0'.63$,
$0'.59\times0'.91\times0'.91$, and $1'.01\times1'.41\times1'.41$,
respectively. } \label{l:xray}
\end{deluxetable}
\end{center}

\end{document}